\documentclass{moriond}
\usepackage{amsmath}

\def\be{\begin{equation}}
\def\ee{\end{equation}}
\def\bea{\begin{eqnarray}}
\def\eea{\end{eqnarray}}

\newcommand{\Photo}{}

\begin{document}
\vspace*{4cm}

\title{ESME stays positive: event generation with positive weights}

\author{M.~van Beekveld}

\address{Nikhef, Amsterdam, The Netherlands}

\maketitle\abstracts{Fully differential simulations of hard-scattering processes are a cornerstone of precision phenomenology at collider experiments. For long, Monte Carlo event generators have been able to achieve next-to-leading order (NLO) accuracy, but their results typically suffer from the presence of negative-weight events. These degrade statistical power and complicate downstream analyses. In this contribution, we present Exponentiated Subtraction for Matching Events (ESME), a novel approach to NLO+PS matching that yields positive-definite event weights while preserving perturbative accuracy and the logarithmic structure of the shower. We discuss the formalism and its implementation within the PanScales framework.}

\section{Introduction}

The precise simulation of high-energy particle collisions requires a consistent description of both the short-distance hard interaction and the subsequent pattern of QCD radiation. This can be achieved through the combination of fixed-order perturbative calculations with parton shower algorithms. General-purpose event generators provide such simulations in the form of fully differential `hard' events, which can subsequently be processed by shower algorithms and hadronisation models.

Over the past decades, significant progress has been made in improving the perturbative accuracy of these simulations. On the resummation side, showers with next-to-next-to-leading logarithmic (NLL) accuracy are becoming the standard~\cite{Dasgupta:2020fwr,Nagy:2020dvz,Forshaw:2020wrq,Herren:2022jej,vanBeekveld:2022ukn,vanBeekveld:2022zhl,vanBeekveld:2023chs,Hoche:2024dee,Preuss:2024vyu}, and crucial steps towards achieving next-to-NLL are being made as well~\cite{vanBeekveld:2023lsa,vanBeekveld:2024qxs,FerrarioRavasio:2023kyg,vanBeekveld:2024wws}. On the fixed-order side, matching schemes such as MC@NLO and POWHEG allow the combination of next-to-leading order (NLO) calculations with parton showers~\cite{Bierlich:2022pfr,Bellm:2015jjp,Sherpa:2024mfk,Alioli:2010xd,Jezo:2015aia,Alwall:2014hca,Frederix:2018nkq}, while more recent developments extend this to next-to-next-to-leading order (NNLO)~\cite{Alioli:2012fc,Alioli:2021qbf,Alioli:2022dkj,Monni:2019whf,Lombardi:2020wju,Mazzitelli:2020jio,Biello:2024pgo}. A persistent issue affecting (N)NLO+PS approaches is the presence of negative-weight events. These arise as a consequence of subtraction procedures and matching prescriptions, and pose both conceptual and practical challenges.

\section{The problem of negative weights}

\begin{figure}[t!]
\centering
  \includegraphics[width=0.5\textwidth]{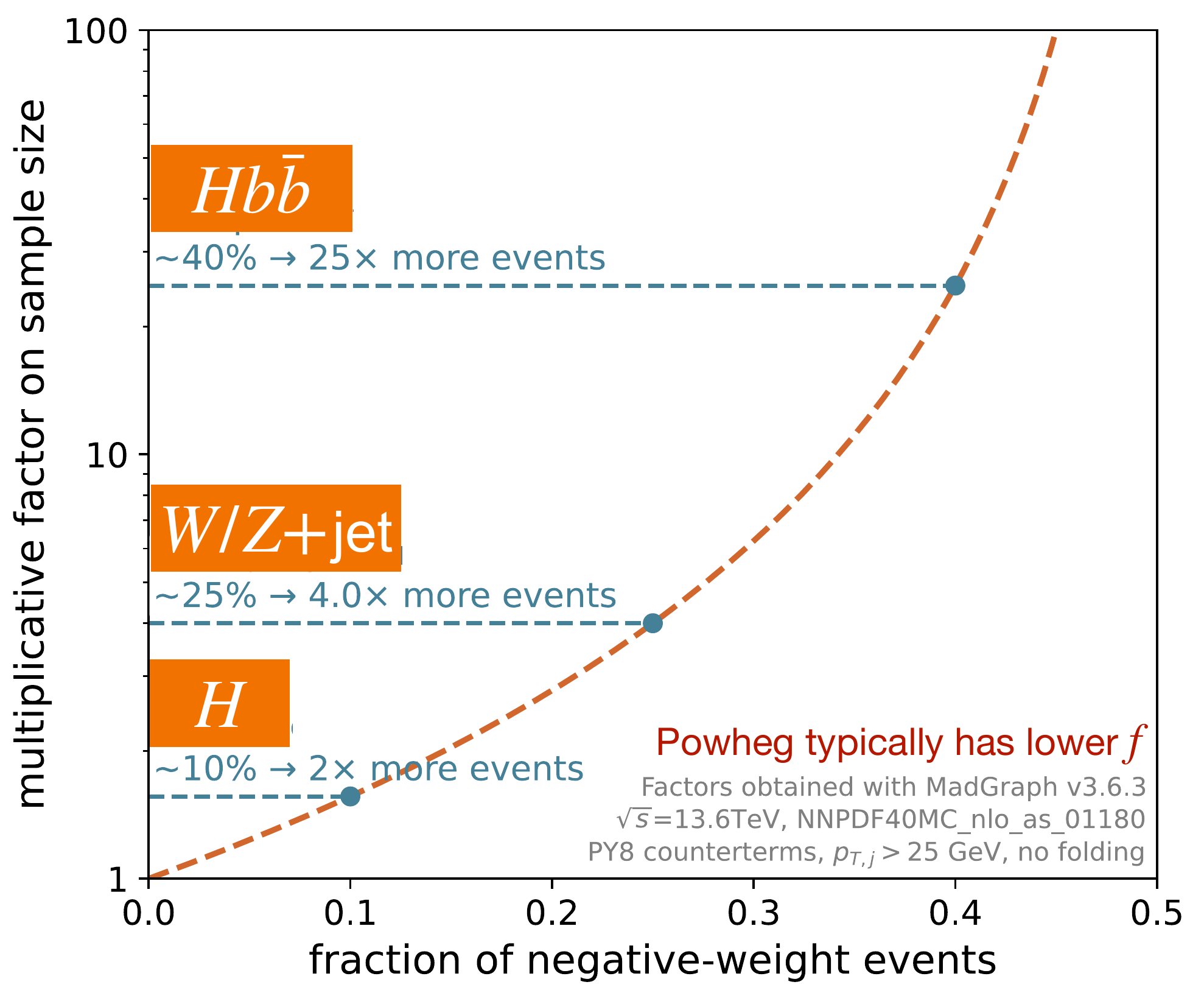}
  \caption{Eq.~\eqref{eq:neg-weights} as a function of $f$, the fraction of negative-weight events. Negative-weight fractions as obtained with MadGraph v3.6.3 for some selected $pp$ collider processes are indicated as well.}
  \label{fig:neg-weights}
\end{figure}

The presence of negative-weight events reduces the statistical efficiency of Monte Carlo simulations. If a fraction $f$ of events carries a negative weight, the effective number of events required to reach a given statistical precision increases according to
\begin{equation}
\label{eq:neg-weights}
N_{\mathrm{eff}} = \frac{1}{(1 - 2f)^2} N \, .
\end{equation}
This implies a substantial increase in computational cost when $f$ is sizeable (see Fig.~\ref{fig:neg-weights} for negative-weight fractions found in typical collider processes). 

Beyond higher computational costs, negative weights also complicate the use of modern analysis techniques, including machine learning methods that rely on probabilistic interpretations of event samples (see e.g.\ Refs.~\cite{ATLAS:2022rws,CMS:2022psv,CMS:2024jdl}). Negative weights are also profoundly unphysical. For these reasons, the theory community has been devoting considerable effort to reduce or eliminate negative weights in event generation (see e.g.\ Refs.~\cite{Jadach:2015mza,Nachman:2020fff,Nason:2021xke,Andersen:2021mvw,Andersen:2023cku,Frederix:2023hom,Shyamsundar:2025nzn,Nachman:2025lid}). 

\section{Structure of NLO+PS event generation}

At NLO accuracy, event generation proceeds by constructing a modified Born-level contribution that incorporates virtual corrections and the integral over real emission contributions. This results in a  modified Born weight
\begin{equation}
\bar{B}(\Phi_B) = B_0(\Phi_B) + (V + C_{\mathrm{int}})(\Phi_B) + \int (R - C)(\Phi_{B+1})\, d\Phi_{\mathrm{rad}} \, ,
\end{equation}
where $B_0$ denotes the squared Born matrix element, $V$ the virtual contribution, $R$ the real-emission matrix element, and $C$ ($C_{\rm int}$) the (integrated) subtraction term. These contributions depend on the phase space $\Phi_B$ and $\Phi_{B+1}$, which describe the underlying Born configuration and the Born plus radiation phase space. The $R-C$ contribution is integrated over the radiation phase space, $\Phi_{\rm rad}$. 

Negative weights can originate from several sources.%
\footnote{Here we will only consider sources of negative weights that are present in a multiplicative matching scheme, such as POWHEG~\cite{Nason:2004rx,Frixione:2007vw}. In additive matching schemes, such as MC\@NLO~\cite{Frixione:2002ik}, additional sources are present.}
First of all, the $\mathcal{O}(\alpha_s)$ correction in $\bar{B}$ can be genuinely large and negative, which happens e.g.\ when one tries to compute a fixed-order correction in a region of phase-space where resummation is important. 
Another source comes from the way the integral for $\bar{B}$ is actually being computed in Monte Carlo algorithms. 
Since it is generally expensive to evaluate the full integral for each Born event fully, often only a (few) single call(s) are made for each Born event (whereas the complete integral in this case can be obtained by simply generating many events).
Of course, positivity of the \emph{integrand} $R-C$ is not guaranteed, as one may happen to sample a region of phase space where $C > R$, rendering to total $\tilde{B}$ (i.e.\ $\bar{B}$ differential in both the born and radiation phase space) negative.
It is precisely this issue that the ESME method aims to tackle. 

\section{Exponentiated Subtraction for Matching Events}
\label{sec:basic-algo}
\begin{figure}[t!]
\centering
  \includegraphics[width=\textwidth]{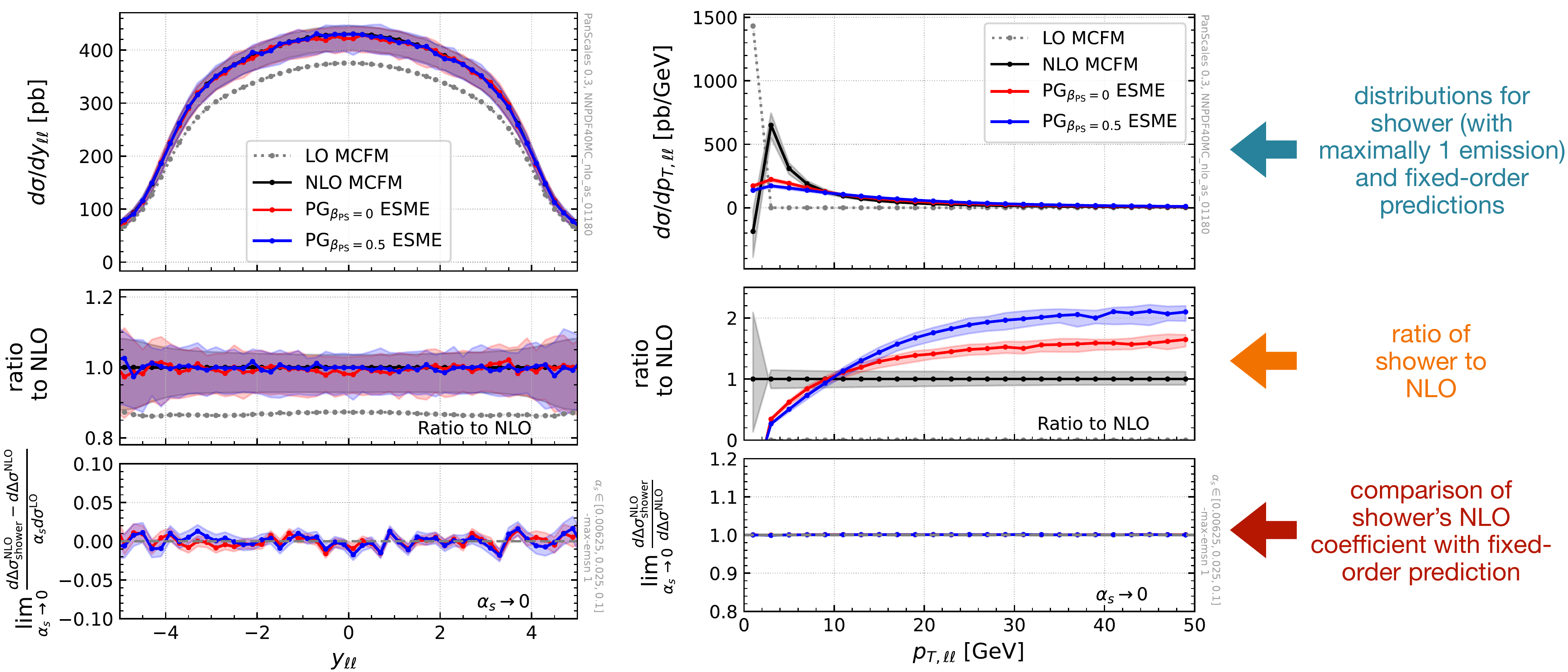}
  \caption{NLO test of the ESME matching method for the $pp \to Z/\gamma^* \to e^+e^-$ process, with the rapidity of the lepton pair on the left, and their transverse momentum on the right.}
  \label{fig:results}
\end{figure}
The Exponentiated Subtraction for Matching Events (ESME) method provides a new way for NLO event generation that avoids negative weights altogether~\cite{vanBeekveld:2025lpz}. The idea relies on expressing $\bar{B}$ as the expectation value of a stochastic variable, allowing the removal of negative weights at the cost of introducing higher-order corrections. 
One ingredient of the ESME method is an algorithm for Exponentiated Subtraction (ES). 
This algorithm is constructed so that it assigns a Born weight $n_B$ for each Born configuration whose average becomes
\begin{equation}
\label{eq:what-we-want}
\langle n_B \rangle = \bar{B}(\Phi_B) + \mathcal{O}(\alpha_s^n)\, .
\end{equation}
The ES algorithm proceeds by introducing an overestimate of the real-emission contribution and iteratively updating the weight $n_B$ through a sequence of probabilistic steps. Starting from an initial value $n_B = 1$, one samples radiation variables and applies accept-reject criteria based on the difference $R - C$. Depending on the sign of this difference, the value of $n_B$ is incremented or decremented by a value of $1/n$. The procedure is designed such that the expectation value of $n_B$ reproduces the NLO-correct weight (see Ref.~\cite{vanBeekveld:2025lpz}, Sec.\ 3.2 for more details). 

In its basic form, the stochastic construction can yield negative values of $n_B$. Such configurations can be removed by rejecting events with $n_B < 0$, thereby ensuring that all retained events carry a positive weight. 
This introduces the $\mathcal{O}(\alpha_s^n)$ correction in Eq.~\eqref{eq:what-we-want}.%
\footnote{Note that this is not unique to the ES algorithm, as any multiplicative matching method will introduce higher-order corrections.}
A key feature of the algorithm is that the order at which these higher-order corrections are introduced are controllable. 
As a result, the method preserves NLO accuracy while achieving strictly positive event weights.

The ES algorithm can be adapted to not only ensure NLO accuracy with positive-weights, but also combine it with real-emission generation and shower evolution, at which point it becomes the ESME algorithm (see Sec.\ 3.3 of Ref.~\cite{vanBeekveld:2025lpz}). 
An important feature of ESME is that it preserves the logarithmic accuracy of the parton shower without requiring additional modifications. This is in contrast to existing HEG-based approaches, where the hardest emission is generated using another phase-space coverage than that of the shower, and consequently care must be taken to preserve the resummation accuracy of the shower~\cite{Hamilton:2023dwb}.

\section{Implementation and results}

The ESME method has been implemented in the PanScales framework~\cite{vanBeekveld:2023ivn} for a range of processes, including $pp\to H/W/Z$ production, deep-inelastic scattering, and $e^+e^- \to q\bar{q}/gg$ scatterings, achieving NLO+NLL accuracy.
Comparisons with fixed-order predictions demonstrate that the method correctly reproduces NLO results for a variety of observables. For example, considering Drell-Yan production, the $Z$-boson rapidity and transverse-momentum distributions agree with those obtained from MCFM~\cite{Boughezal:2016wmq} (see Fig.~\ref{fig:results}).
From a computational perspective, the method offers a significant improvement in efficiency. In general, the absence of negative weights reduces the number of events required to reach a given statistical precision. The precise speed-up depends, of course, on the fraction $f$ of negative weights present in the original setup, which is relatively small for the colour-singlet processes considered here.
Independently of this, the implementation of ESME in the PanScales framework achieves speed-ups of up to an order of magnitude compared to existing NLO tools when comparing the event generation time spent per event (see Fig.~12 of Ref.~\cite{vanBeekveld:2025lpz}).
This demonstrates that eliminating negative weights does not require compromising computational speed as a trade-off.

\section{Conclusions}

We have presented a new approach to NLO+PS matching based on exponentiated subtraction, which enables the generation of positive-weight events while preserving the logarithmic accuracy of parton showers. The method has been validated for a range of processes and demonstrates both theoretical robustness and practical efficiency.

Future developments will focus on extending the method beyond colour-singlet production and towards higher perturbative orders. These directions are expected to both speed up the generation and enhance the precision of Monte Carlo event generators.

\section*{Acknowledgments}

MvB thanks G.\ P.\ Salam and L.\ Scyboz for comments on the manuscript.
The results presented in these
proceedings have been obtained in collaboration with S.\ Ferrario Ravasio, J.\ Helliwell, A.\ Karlberg, G.\ P.\ Salam, L.\ Scyboz, A.\ Soto Ontoso, G.\ Soyez and S.\ Zanoli.
This work was supported by the Dutch Research Council (NWO) under project number VI.Veni.232.190. 
AI tools were used to assist drafting this document. 

\bibliography{mvb_moriond}


\end{document}